\font\sym=lasy10

\def\qed{{\sym \char50}}
\def\frac#1#2{{#1\over#2}}

\def\ip#1#2{\langle #1,#2\rangle}
\def\expc#1{\langle #1\rangle}
\def\conj#1{\overline{#1}}
\def\tr{{\rm tr}}
\def\hsp{{\cal H}}
\def\alg{{\cal A}}
\def\car{{\rm CAR}}
\def\vect{{\rm vect}}
\def\Diff{{\rm Diff}}
\def\real{{\bf R}}
\def\complex{{\bf C}}
\def\integer{{\bf Z}}
\def\natno{{\bf N}}
\def\im#1{\widetilde{#1}}
\def\pd#1#2{\frac{\partial #1}{\partial #2}}
\def\sec#1{\advance\count31 by 1\count34=0\bigskip\noindent
{\bf \the\count31. #1}\medskip}
\def\app#1{\advance\count33 by 1\bigskip\noindent{\bf Appendix
\the\count33. #1}
\medskip}
\def\ref#1{\advance\count32 by 1\item{\the\count32.}}
\def\thm#1{\advance\count34 by 1\bigskip
{\bf #1 \the\count31.\the\count34.}}
\def\pf{\medskip{\it Proof. }}
\def\endpf{\hfill\qed\bigskip}

\count31=0
\count32=0
\count33=0
\count34=0

\noindent{\bf Boundary conformal fields and Tomita--Takesaki theory}

\bigskip
{\narrower
\noindent{K.C.\ Hannabuss and M.\ Semplice$^*$}
\medskip\noindent
{\it Mathematical Institute, 24-29, St Giles', Oxford OX1 3LB, England}
\smallskip
\noindent{\it $^*$current address: Dipartimento di Matematica, Via
Saldini 50, 20133 Milano, Italy}
\smallskip}

\bigskip
{\narrower
\noindent
Motivated by formal similarities between the continuum limit of the Ising
model and the Unruh effect, this paper connects the notion of an Ishibashi
state in boundary conformal field theory with the Tomita--Takesaki theory
for operator algebras.
A geometrical approach to the definition of Ishibashi states is presented,
and it is shown that, when normalisable the Ishibashi states are cyclic
separating states, justifying the operator state corespondence.
When the states are not normalisable Tomita--Takesaki theory offers an
alternative approach based on left Hilbert algebras, opening the way to
extensions of our construction and the state-operator correspondence.
\smallskip}

\sec{Introduction}

Since their introduction and exploitation, particularly by Cardy, there
has been a strong interest in boundary states in conformal field theory
[3,4,5,6,14].
However, there are many other interesting examples of quantum field
theories with boundary.
For example, the Unruh effect [25] in which an observer accelerating
through a vacuum detects a thermal spectrum of particles, can be linked to
the splitting of two-dimensional Minkowski space into two Rindler-type
space-times, and the horizon or boundary between them (see particularly
[22,25,2]).
An even more obvious example of a boundary, though in momentum rather than
configuration space, is the Fermi energy level.
Although physically very different, these share mathematical features
which we shall study in this paper, placing boundary conformal field
theory within the broader context of operator algebras associated to
quantum field theories with boundaries.
We shall concentrate on the boundary states, where the broader context
suggests an alternative mathematical description of Ishibashi states,
which avoids the normalisability problem.
The key mathematical tool, suggested already by the Unruh effect, is
Tomita--Takesaki theory, whose primary physical use is usually to study
thermal states and the KMS condition.

To see how this comes about we first consider the treatment of the
Ishibashi boundary states in conformal field theory.
In the physics literature conformal symmetry is usually expressed in terms
of a Lie algebra which is the direct sum of two (commuting) copies of the
Virasoro algebra, a central extension of the vector fields $\vect(S^1)$ on
a circle.
The Virasoro algebra is generated by elements $L_n$ (or by $\im{L}_n$ for
the other copy) for $n\in \integer$, with the Lie brackets given in terms
of the central charge $c\in \real$ by
$$[L_m,L_n] = (m-n)L_{m+n} + \frac{cn(n^2-1)}{12} \delta_{m+n,0}, \qquad
[\im{L}_m,\im{L}_n] = (m-n)\im{L}_{m+n} + \frac{cn(n^2-1)}{12}
\delta_{m+n,0}.$$
These commutation relations define projective representations $L$ and
$\im{L}$ of $\vect(S^1)$ with essentially the same multiplier when we
identify
$L_n = -L(z^{n+1}\partial/\partial z)$ and $\im{L}_n
=-\im{L}(\conj{z}^{n+1}\partial/\partial\conj{z})$.
We shall often write $L(X)$ for the representation of a general
holomorphic vector field $X$, and $\im{L}(Y)$ for an
antiholomorphic vector field $Y$, and use $\sigma$ for the multiplier.

The Ishibashi states $\Omega$ are supposed to satisfy
$L_n^*\Omega = \im{L}_n\Omega$, for all $n\in \integer$,
(where $L_n^* = L_{-n}$).
This condition can be regarded as a replacement for the highest weight
condition that $L_n\Omega = 0 = \im{L}_n\Omega$ for $n>0$.
Unfortunately, the vectors $\Omega$ which occur in the physics literature
are
almost always unnormalisable, that is, they are not really vectors in the
representation space at all.

Another useful feature of conformal field theories is the operator--state
correspondence in which an algebra element $a$ is identified
with the vector $a\Omega$.
In fact, this identification map is surjective just when the vector
$\Omega$ is a cyclic vector, and is one-one when $\Omega$ is separating
(that is, $a\Omega$ vanishes only when $a=0$).
The operator-state correspondence also means that the algebra can also be
regarded as an inner product space, and so should in fact be some sort of
Hilbert algebra (a $*$-algebra which is also a pre-Hilbert space with
certain properties linking the multiplication and inner product).
Now left Hilbert algebras and cyclic separating vectors are united in
Tomita--Takesaki theory, [24,7,8,22].
Moreover, that theory can cope with the situation when there is only the
Hilbert algebra, but no cyclic separating vector $\Omega$, as happens when
the Ishibashi states are not normalisable.

In this setting other features of conformal field theory find a natural
place.
For example, left and right multiplication in the Hilbert algebra generate
two commuting von Neumann algebras of operators, which Tomita theory shows
to be anti-isomorphic.
This is just the sort of structure exhibited by the two algebras generated
by the $L_n$ and the $\im{L}_n$.
In fact, we shall often find it convenient to forget the detailed
structure of the Virasoro algebra and simply work with two commuting (or
graded commuting) algebras $\alg_+$ and ${\alg_-}$ which are related by
some conjugate linear homomorphism $a \mapsto \im{a}$, such that
$\im{\im{a}} = a$.
We associate to the boundary a left Hilbert algebra, $\alg_0$, having the
given commuting anti-isomorphic algebras as its left and right von Neumann
algebras.
In some cases this Hilbert algebra can be generated by a generalised
Ishibashi vector $\Omega$ satisfying $a^*\Omega = \im{a}\Omega$, for all
$a\in \alg_+$.

In this paper we shall show how this viewpoint enables us to reconstruct
various results from the physics literature.
Section 2 explains how the geometrical link between boundaries and
involutions provides an easy characterisation of a subgroup of the
conformal group respecting the boundary, and of the Ishibashi states.
This is extended in Section 3 to $*$-algebras having an antilinear
involutory automorphism.
In Section 4 it is shown that these definitions pick out cyclic separating
vectors, bringing the ideas into the framework of Tomita--Takesaki theory.
Section 5 looks at properties of symmetries of such a system.
The ideas are brought together in Section 6 to show how Tomita--Takesaki
theory provides a replacement for Ishibashi states when, as usually
happens, these are not normalisable.
Finally we discuss the situation when a region has several disconnected
boundaries.
The two appendices review the Unruh effect from the perspective of
conformal field theory and the fermion field theory arising as the
continuum limit of the Ising model.

Whilst working on this topic we became aware of work by Wassermann [27]
which also investigates boundary conformal field theory using operator
algebras, but with somewhat different objectives.
The monograph by Evans and Kawahigashi explains the links between operator
algebras and ordinary conformal field theory.

\sec{The conformal group}

The vector fields are the Lie algebra of the orientation preserving
diffeomorphisms of the circle  $H = \Diff^+(S^1)$.
In practice, however, the Lie algebra action of vector fields does not
always exponentiate to a well-defined action of $H$, and, as Isham has
remarked [13], it really makes more sense to consider a pseudogroup of
locally defined transformations.
Alternatively, one might allow for groups or Lie algebras by working in
the context of a Hopf algebra, but, for simplicity, having signalled the
technical obstacle, we shall proceed as though the group actions existed,
leaving the reader to reinterpret results in those few cases where
necessary.

The key to the study of boundaries in quantum conformal theory, as in its
classical analogue, is the method of images.
The boundary separates two regions, the physically interior region and its
reflection outside the curve.
The reflection, which reverses the holomorphic structure, fixes the
boundary.
For example, in two-dimensions the unit circle $C$ is the fixed point set
of the antiholomorphic involution $\kappa_C:z\mapsto 1/\conj{z}$ which
interchanges the unit disc and the exterior, whilst the real axis is the
set fixed by conjugation $\kappa_\real: z\mapsto \conj{z}$.
By the Riemann mapping theorem the interior of any Jordan curve in
$\complex$ can be mapped to the unit disc by a map $\Phi$, so for any
such curve there is an antiholomorphic involution $\kappa =
\Phi^{-1}\kappa_C\Phi$ which interchanges the inside and
outside of the curve, (though one has to be careful about behaviour on the
curve itself).
(In practice, it is more convenient to use the map $F$ taking the interior
to the upper half plane, and $\kappa = F^{-1}\kappa_\real F$, so that
$\kappa(z) = F^{-1}\conj{F(z)}$, where $\conj{F}(z) =
\conj{F(\conj{z})}$.)
The product of two antiholomorphic involutions is holomorphic
(for example, $\kappa\kappa_\real(z) = F^{-1}\conj{F}(z)$),
and so products of even numbers of such involutions generate a
subgroup of the conformal group, which is clearly normal as the conjugate
of a product of involutions is the product of their conjugates.
Using the fact that the conformal group is the product of two copies of
the diffeomorphism group of the circle, together with Cartan's result
that diffeomorphism groups have simple Lie algebras [12], we see that a group
with the Lie algebra of the whole conformal group is generated in this
way.

The boundary involutions induce antilinear automorphisms of any algebras
associated to the surface, and we shall argue that these provide a dense
subalgebra with the structure of a Tomita or modular Hilbert algebra,
which encodes the information about the boundary normally described using
Ishibashi states.

A conformal transformation of $S$ can be reflected to give a conformal
transformation in the subgroup $G_\kappa$ commuting with the involution
$\kappa$.

\thm{Lemma}
The restriction of the multiplier to $G_\kappa$ is trivial.

\pf
We start by considering the case of the upper half plane and involution
$\kappa = \kappa_\real:z\mapsto \conj{z}$ defining the real axis.
A conformal transformation $F: z \mapsto F(z)$ commutes with
$\kappa_\real$ if and only if $F(\conj{z}) = \conj{F(z)}$, or
equivalently $F = \conj{F}$.
To find the effect on the multiplier we need to work at the Lie algebra
level, where a typical vector field has the form
$$X+Y =
\sum_{n} X_nz^{n+1}\pd{}{z} + \sum_{n} Y_n\conj{z}^{n+1}\pd{}{\conj{z}}.$$
We easily calculate that
$$\kappa(X+Y)\kappa =
\sum_{n} \conj{Y}_nz^{n+1}\pd{}{z}
+ \sum_{n} \conj{X}_n\conj{z}^{n+1}\pd{}{\conj{z}},$$
so that $X+Y$ commutes with $\kappa$ if and only if $Y_n = \conj{X}_n$ for
all $n$, (or equivalently $Y = \kappa X\kappa$).
In the real Lie algebra of $G$ we also have $X_{n} = -\conj{X}_{-n}$.
Thus in the real Lie algebra of $G_\kappa$ one has $X_{n} = -Y_{-n}$,
so that it is generated by elements of the form
$z^{-n+1}\partial/\partial{z} -
\conj{z}^{n+1}\partial/\partial{\conj{z}}$.
(In more abstract form the elements of this subalgebra have the form
$X+X^\kappa$, where $X^\kappa = \kappa X\kappa$.)

The representations are thus generated by $L_{-n}-\im{L}_{n}$.
Now, since the $L_m$ and $\im{L}_n$ commute,
$$[L_{-m}-\im{L}_m,L_{-n}-\im{L}_n] + (m-n)(L_{-m-n} - \im{L}_{m+n})
= [L_{-m},L_{-n}] + (m-n)L_{-m-n} + [\im{L}_m,\im{L}_n] -
(m-n)\im{L}_{m+n}.$$
The first two terms give $-\delta_{m+n,0}cn(n^2-1)/12$, whilst the last
pair gives the same with $n$ replaced by $-n$, so that there is
cancellation, and the multiplier vanishes on this subalgebra.

Although we have only proved the result for $\kappa_\real$,  any other
involution is conjugate to this and conjugation does not affect the
triviality of the multiplier.
\endpf

\bigskip
{\it Note.}
The characterisation of the elements of the Lie subalgebra as having the
form $X+X^\kappa$, works more generally, and these are represented by
$L(X)+\im{L}(X^\kappa)$.
For unitary representations of the real algebra $L(X) = -L(X)^*$, so that
the subalgebra is represented by elements of the form
$-L(X)^*+\im{L}(X^\kappa)$.
One then checks that, for any holomorphic vector fields $X$ and $Y$
$$\eqalign{[-L(X)^*+\im{L}(X^\kappa),-L(Y)^*+\im{L}(Y^\kappa)]
&+L([X,Y])^* -\im{L}([X,Y]^\kappa])\cr
&= [L(X)^*,L(Y)^*] + L([X,Y])^* + [\im{L}(X^\kappa),\im{L}(Y^\kappa)]
-\im{L}([X^\kappa,Y^\kappa])\cr
&= -\left([L(X),L(Y)] - L([X,Y])\right)^* +
\left([\im{L}(X^\kappa),\im{L}(Y^\kappa)] -
\im{L}([X^\kappa,Y^\kappa])\right),\cr}$$
which cancels to give 0.

\bigskip
The corresponding condition for the conformal group $G=H\times H$ is
obtained by taking the tensor product ${\cal V}$ of the
$\sigma$-representations $V$ and
$\im{V}$ obtained by exponentiating $L$ and $\im{L}$: $V(\exp(X)) =
\exp(L(X))$ and $\im{V}(\exp(X)) = \exp(\im{L}(X))$.
As we readily see, the subgroup commuting with the involution is
$$G_\kappa = \{(x,x^\kappa)\in H\times H: x\in H\}.$$
We may look for a vector $\Omega_\kappa$ in the
representation space which is an eigenvector for all elements $g\in
G_\kappa$:
$${\cal V}(g)\Omega_\kappa = \lambda(g)\Omega_\kappa.$$
This is a quantum mechanical analogue of the curve itself for a conformal
field theory based in the interior of the fixed point set of $\kappa$.
(For consistency the multiplier on the subgroup $G_\kappa$ must be
trivial, but that is assured by the Lemma.)

The eigenvector $\Omega_\kappa$ must also be an eigenvector for the Lie
algebra of $G_\kappa$ and, when the boundary is the real axis, we know
that this is generated by $L_{-n}-\im{L}_{n} = L_n^*-\im{L}_n$.
The simplest case is when the eigenvalues vanish, (or $\Omega_\kappa$ is
actually fixed by the subgroup $G_\kappa$) giving
$(L_n^*-\im{L}_n)\Omega_\kappa = 0$, for all $n\in \natno$, which is the
Ishibashi condition.
This condition can also be expressed in the form
$L(X)^*\Omega_\kappa = \im{L}(X^\kappa)\Omega_\kappa$, valid for any
boundary curve.
(When the eigenvalue is non-vanishing one may subtract half of it from
each $L$ and $\im{L}$, to obtain new operators satisfying the same
commutation relations whose kernel contains $\Omega_\kappa$, so that the
condition that the vector be fixed by $G_\kappa$ is less special than
appears at first sight.)
We deduce the following result.

\thm{Lemma}
The Ishibashi condition on a vector $\Omega_\kappa$ is equivalent to
$\Omega_\kappa$ being a vector fixed the representation of $G_\kappa$, or
annihilated by its Lie algebra.

\bigskip
{\it Note:}
It follows from the definition of the Ishibashi boundary state $\Omega$
that
$$\ip{\Omega}{\im{V}(\im{x}_j)V(x_k)\Omega}
=\ip{\im{V}(\im{x}_j^{-1})\Omega}{V(x_k)\Omega}
=\ip{V(x_j^{-1})^*\Omega}{V(x_k)\Omega}
=\ip{V(x_j)\Omega}{V(x_k)\Omega}$$
defines a positive matrix.
In Euclidean algebraic field theory this is the reflection positivity
condition, [18,11].

The advantage of this more abstract characterisation is that similar
constructions could be made for any group $G$ with a multiplier $\sigma$
with subgroups $H$, on which the multiplier is totally non-degenerate,
and $K$ on which $\sigma$ is trivial, such that $H\cap K = \{1\}$ and
$G = HK$.
In some ways the special feature of conformal field theory is that all
boundaries are (more or less) equivalent.
The mass $m$ bosons in the positive $z$ half of $\real^3$ with Dirichlet
boundary conditions, for example, still have an obvious Green's function
$$G_\real({\bf r},{\bf a})
= \frac{e^{-m|{\bf r}-{\bf a}|}}{4\pi|{\bf r}-{\bf a}|}
- \frac{e^{-m|{\bf r}-\im{{\bf a}}|}}{4\pi|{\bf r}-\im{{\bf a}}|},$$
where $\im{{\bf a}}$ is the reflection of ${\bf a}$ in the plane $z=0$.
However, the same Dirichlet problem in the unit sphere has Green's
function
$$G_C({\bf r},{\bf a})
= \frac{e^{-m|{\bf r}-{\bf a}|}}{4\pi|{\bf r}-{\bf a}|} -
\frac{e^{-m\lambda|{\bf r}-\im{{\bf a}}|}}{4\pi\lambda|{\bf r}-\im{{\bf
a}}|},$$
where $\im{{\bf a}}$ is now the inverse of ${\bf a}$ with respect to the
sphere, and $\lambda = 1/|{\bf a}|$, so that this non-conformally
invariant system has rather different forms of Green's function for the two
boundaries.

\sec{Boundary states for algebras}

We may encode the effect of the boundary on the conformal Lie algebra by
defining the map
$$\alpha_\kappa[L(X) + \im{L}(Y)] = L(X^\kappa)+\im{L}(Y^\kappa).$$
From the properties of $\kappa$ it is clear that $\alpha_\kappa$ is an
antilinear involution, and, using the same argument as in the alternative
proof of Lemma 2.1, $\alpha_\kappa$ is an additive and multiplicative
$*$-homomorphism.
It is therefore an antilinear automorphism.

Returning to the general situation, we write $\alg_+$ for the algebra of
fields in $S$, ${\alg_-}$ for those on $\im{S}$, and $\Omega$ for the
boundary state in a space on which both algebras operate.
We assume that an involutory antilinear $*$-isomorphism
$\alpha_\kappa:{\alg_+}\to {\alg_-}$ can be associated with the geometric
involution $\kappa$.
We shall sometimes write $\alpha_\kappa(a) = \im{a}$.
This can be extended to an involution $\alpha_\kappa$ of the algebras
generated by $\alg_+$ and ${\alg_-}$ by defining
$\alpha_\kappa|_{\alg_-} = \alpha_\kappa^{-1}|_{\alg_-}$.
A boundary state $\Omega_\kappa$ is required to satisfy
$$a^*\Omega_\kappa = \alpha_\kappa(a)\Omega_\kappa,$$
for all $a\in \alg_+$, and since $\alpha_\kappa$ is an involution,
the same applies to the whole algebra generated by ${\alg_+}$, and
${\alg_-}$.

We have seen that these relations hold when ${\alg_+}$ is the enveloping
algebra of one copy of the Virasoro algebra and ${\alg_-}$ the other, or
when $\alg_+$ and ${\alg_-}$ are suitable group algebras for the
corresponding groups.
However, there are other examples such as the massless free fermion theory
which is the continuum limit of the Ising model [10,16], (see Appendix 2).
Fermion theories are described by canonical anticommutation relation
algebras CAR($W$) over a complex inner product space $W$, and are generated by
creation operators $c(w)$, depending linearly on $w\in W$, and their
adjoint annihilation operators, satisfying the canonical anticommutation
relations
$$[c(w)^*,c(z)]_+ = \ip{w}{z}1, \qquad [c(w),c(z)]_+ = 0.$$
The papers [10,16] describe the boundary states in terms of a Bogoliubov
transformation $K$.
This would normally be given in terms of Bogoliubov operators $A$ (linear)
and $B$ (antilinear) on $W$, which would be thought of as defining an
automorphism of the CAR algebra:
$$c_{(A,B)}(w) = c(Aw)-c(Bw)^*.$$
The conditions for this to be an automorphism ($c_{(A,B)}$ and $c$ satisfy
the same anticommutation relations) can be written as
$$A^*A+B^*B = 1, \qquad A^*B+B^*A = 0.$$
When $A$ is invertible we may introduce the antilinear operator $Z=
BA^{-1}$ and rewrite the second condition as $Z+Z^* = 0$.
The connection with the Ishibashi states comes from the observation that
the condition $c_{(A,B)}(w)^*\Omega=0$ (for all $w\in W$) defining a Fock
vacuum $\Omega$, can be rewritten as
$$c(w)^*\Omega  = c(Zw)\Omega,$$
which looks like an Ishibashi condition with $\alpha_\kappa(c(w)) =
c(Zw)$.

The problem with this approach is that in the case of the Ising model $Z$
is not a Hilbert--Schmidt operator, and so (by the Shale--Stinespring
criterion, [20]) the Bogoliubov transformation is not implementable, as
the papers acknowledge, so that $\Omega$ does not lie in the same
representation space as the Fock vacuum for $c$.
However, in this case the space $W$ decomposes into $W_+\oplus W_-$, the
orthogonal direct sum of two subspaces, corresponding to the two sides of
the boundary and the Ishibashi criterion is needed not for all $w\in W$, but
only for $w$ in the subspace $W_+$.
This provides an alternative interpretation of the condition on $\Omega$.

Suppose that (as happens in the example) $A$ maps each of $W_\pm$ to
itself, whilst $B$ sends $W_\pm$ to $W_\mp$.
The condition that $A^*B+B^*A$ should vanish is now automatically
satisfied on the subspace $W_+$, though the condition $A^*A+B^*B = 1$ is
still needed.
We shall write $c_\pm$ for the restriction of $c$ to $W_\pm$, and then we
have $c_{(A,B)}(w) = c_+(Aw)-c_-(Bw)^*$, for $w\in W_+$.
This formula is essentially the Araki--Powers--St\o rmer purification map,
[1,18], which realises a quasi-free state of $W_+$ as the restriction
of a Fock state for the \lq\lq doubled\rq\rq\ space $W = W_+\oplus W_-$.
(Quasi-free states have all their $n$-point correlation functions given in
terms of the 2-point correlation functions by the same formulae as for
Fock states, for example in the fermion case by Wick's determinant
formula.)
Purification is generally used when $Z$ is invertible (so that $W$ really
is a double), and $Z$ need no longer satisfy a Hilbert--Schmidt condition.
In the example of the Ising model $Z$ is indeed invertible, and
this provides a better interpretation of the Ishibashi condition.

Before stating the key result we note that this example shares with the
conformal algebra the property that there are simple commutation relations
between $\alg_+$ and ${\alg_-}$ (which intersect only in $\complex 1$).
For the Ising model $\alg_+ = \car(W_+)$, and ${\alg_-} = \car(W_-)$.
We shall assume that in general we have a relation of the sort
$$a\alpha_\kappa(b) = \epsilon(b,a^*)\alpha_\kappa(b)a,$$
with $\epsilon(b,a^*)\in \complex$, for all $a,b\in \alg_+$.
(In the case of the Virasoro algebra $\epsilon(a,b)$ is identically 1, and
for homogeneous elements of the CAR algebra of degrees $d(a)$ and $d(b)$
it is $(-1)^{d(a)d(b)}$.)
For consistency we now require
$$\alpha_\kappa(ab)\Omega_\kappa = b^*a^*\Omega_\kappa
= b^*\alpha_\kappa(a)\Omega_\kappa
= \epsilon(a,b)\alpha_\kappa(a)b^*\Omega_\kappa
= \epsilon(a,b)\alpha_\kappa(a)\alpha_\kappa(b)\Omega_\kappa,$$
suggesting that $\kappa$ should satisfy $\alpha_\kappa(ab) =
\epsilon(a,b)\alpha_\kappa(a)\alpha_\kappa(b)$.
In practice algebras such as the Virasoro and CAR algebras are graded and
we can use this formula as a way of generating the whole algebra from its
degree one subspace, which is where the condition on $\Omega_\kappa$ is
initially given.

When the algebra and its image enjoy a commutation relation of this sort
they generate the algebra $\alg = \alg_+\alpha_\kappa(\alg_+)$.
One can, if so desired, generalise the notion of crossed product to this
setting and work with the crossed product $\expc{\kappa}\bowtie\alg$ of
the algebra $\alg_+\alpha_\kappa(\alg_+)$ by the group $\expc{\kappa}
\cong \integer_2$ generated by $\kappa$.

\sec{Tomita--Takesaki theory}

We now turn to a very important property of $\Omega_\kappa$, which does
not seem to have been given much prominence.
In the presence of a boundary the algebras are doubled due to reflection,
and we have seen how this doubling can be interpreted as a version of the
Araki--Powers--St\o rmer (APS) purification construction.
(This already links it to numerous quite different physical situations
where quasifree states appear naturally, as, for example, for systems at
non-zero temperatures.)

The cyclic vector of the quasi-free states constructed by non-trivial
doubling is usually also separating, that is $a\Omega=0$ for $a\in\alg_+$
only if $a=0$.
In fact this is easy to prove directly.

\thm{Theorem}
Suppose that $\alg_+$, $\alpha_\kappa$, $\epsilon$ are as above, and
that $\hsp$ is a module for $\alg_+\alpha_\kappa(\alg_+)$.
If there exists a cyclic vector $\Omega_\kappa$ satisfying
$\alpha_\kappa(a)\Omega_\kappa = a^*\Omega_\kappa$ for all $a\in \alg_+$,
then it is cyclic and separating for $\alg_+$.

\pf
The commutation property for $\alg_+$ and $\alpha_\kappa(\alg_+)$ permits
us to order any product of elements of $\alg_+$ and ${\alg_-}$ with the elements
of $\alg_+$ to the left, and those of ${\alg_-}$ to the right.
If $\Omega_\kappa$ is a cyclic vector for the double algebra then the
space is the closure of the span of products acting on $\Omega_\kappa$.
Now, any element of ${\alg_-}$ has the form $\alpha_\kappa(a)$ for
$a\in\alg_+$, and, if $\alpha_\kappa(a)\Omega_\kappa = a^*\Omega_\kappa$
for all $a\in \alg_+$, then this can be replaced by $a^*\Omega_\kappa$.
Using the commutation property $a^*$ can be taken to the left of the other
elements of ${\alg_-}$ acting on $\Omega_\kappa$, and the process repeated
until we have only elements of $\alg_+$ acting on $\Omega_\kappa$, showing
that $\alg_+$ also generates the whole space from $\Omega_\kappa$.
We could have argued similarly that $\Omega_\kappa$ is also cyclic for
${\alg_-}$, which is equivalent to its being separating for $\alg_+$.
(For if $a\Omega_\kappa = 0$ for $a\in \alg_+$, then for any $b\in \alg_+$
we have
$$a\alpha_\kappa(b)\Omega_\kappa
= \epsilon(b,a^*)\alpha_\kappa(b)a\Omega_\kappa = 0,$$
and, since $\Omega_\kappa$ is also cyclic for ${\alg_-}$, this shows that
$a$ annihilates the whole space,  so that $a=0$.)
\endpf

\bigskip
We may now define the Tomita operator
$S_\kappa: a\Omega_\kappa \mapsto a^*\Omega_\kappa$, for $a\in \alg_+$.
By definition $S_\kappa$ is an involution and fixes $\Omega_\kappa$, but
also
$$\eqalign{S_\kappa a S_\kappa b\Omega_\kappa &= S_\kappa
ab^*\Omega_\kappa\cr
&= ba^*\Omega_\kappa\cr
& = b\alpha_\kappa(a)\Omega_\kappa\cr
&= \alpha_\kappa(a)b\Omega_\kappa,\cr}$$
showing that $\alpha_\kappa(a) = S_\kappa aS_\kappa$.
Thus we may obtain an action of the crossed product by sending $\kappa$ to
$S_\kappa$.

We have already noted that a cyclic separating vector is precisely what is
needed to justify the state-operator correspondence, since there is a
one-one correspondence between algebra elements $a\in \alg_+$ and the vectors
$a\Omega_\kappa$.
(This has long been known in quantum field theory in the context of the
Reeh--Schlieder theorem.
A similar connection between cyclic separating vectors and reflection
properties has been used purely as a mathematical tool in [15].)
In Tomita--Takesaki theory this correspondence is used to give the algebra
an inner product $\ip{a}{b} = \ip{a\Omega_\kappa}{b\Omega_\kappa}$ with
respect to which it is a left Hilbert $*$-algebra [24,7,8,23].
(This is a *-algebra, which is also an inner product space, such that the
map $a \mapsto a^*$ is closable, the left multiplication action of the
algebra on itself  defines a bounded non-degenerate $*$-representation.)
In conformal field theory one tends to work with the much smaller algebra
of primary fields. This has the advantage of giving a much smaller
Frobenius algebra, but loses other structure such as the adjoint.

Tomita--Takesaki theory gives us far more than this.
The operator $S_\kappa$, defined above, has a polar decomposition with
positive part given by the positive linear operator $\Delta_\kappa =
S_\kappa^*S_\kappa$, and antiunitary part $J_\kappa
= S_\kappa\Delta_\kappa^{-\frac12}$, which is also an involution.
(The association of boundary states to antiunitary operators has been
noted in a somewhat different form by Watts.
One can construct a representation of the cross product by mapping
$\kappa$ to
$S_\kappa$, but when $\Delta_\kappa \neq 1$ this is not antiunitary, and
so one does not obtain a $*$-representation.)
It is then known that the state defined by $\Omega$ satisfies the KMS
condition at inverse temperature 1, with respect to the one-parameter
unitary automorphism group $a \mapsto a_t=\Delta^{it}a\Delta^{-it}$ that is
$$\ip{\Omega}{ab\Omega} = \ip{\Omega}{b\Delta a\Delta^{-1}\Omega}.$$
It is also known that $J$ defines a spatial anti-isomorphism between
$\alg_+$ and its commutant $\alg_+^\prime$ (the operators on the space
$\alg_+\Omega$ which commute with $\alg_+$), that is
$\alg_+^\prime = J\alg_+J^{-1}$.
(In the Ising model the commutant is a modified version of CAR($W_-$).)
In fact $J$ is also an involution.
In conformal theories it can be considered as representing $\kappa$ in a
unitary-antiunitary representation of the conformal group extended by
$\kappa$.

\sec{Symmetries of the system}

Usually the physical algebra will also have symmetries, acting as
automorphisms, as, for example, the conformal group acts as automorphisms
of the CAR algebra.
We can then form the crossed product of the symmetry group and algebra.
For boundary theories it makes sense to consider a group $G$ which
contains  the symmetry group $G_0$ as a normal subgroup of index 2, where
we think of  $G$ as the extension of $G_0$ by the addition of the boundary
involution $\kappa$.
The group $G_0$ acts by automorphisms $\alpha_g$ of $\alg_+$ and elements
of the non-trivial coset in $G/G_0$ by antilinear automorphisms.
For consistency the map $g\mapsto \alpha_g$ is a homomorphism, which means
that $\alpha_\kappa\alpha_g\alpha_\kappa = \alpha_{\kappa g\kappa}$.

The $*$-representations of the crossed product algebra correspond
naturally to covariant representations $(V,\pi)$ consisting of a
projective representation $V$ of the group and a $*$-representation
$\pi$of the algebra, which satisfy $V(g)\pi(a) = \pi(\alpha_g(a))V(g)$.

\thm{Lemma}
Let $(V,\pi)$ be a covariant representation of $(G,\alg_+)$, with
consistency between the involutions in the sense that
$\alpha_\kappa\alpha_g\alpha_\kappa = \alpha_{\kappa g\kappa}$, and
suppose that there is a unique generalised Ishibashi vector
$\Omega_\kappa$ for $\alg_+$.
Then $\Omega_\kappa$ is also an eigenvector for $G$.

\pf
Using the covariance condition in the form
$V(g)\pi(a)^* = \pi(\alpha_g(a))^*V(g)$, we therefore have
$$\pi(\alpha_g(a))^*V(g)\Omega_\kappa = V(g)\pi(a)^*\Omega_\kappa
= V(g)\pi(\alpha_\kappa(a))\Omega_\kappa
= \pi(\alpha_{g}\alpha_\kappa(a))V(g)\Omega_\kappa.$$
When $g\in G_\kappa$ this can be written as
$\pi(\alpha_\kappa\alpha_{g}(a))V(g)\Omega_\kappa$.
Replacing $\alpha_g(a)$ by $a$ gives
$$\pi(a)^*V(g)\Omega_\kappa
= \pi(\alpha_\kappa(a))V(g)\Omega_\kappa,$$
so that by uniqueness $V(g)\Omega_\kappa$ is a multiple of
$\Omega_\kappa$,
showing that $\Omega_\kappa$ also defines a boundary state for $G$.
\endpf

\sec{Left Hilbert algebras}

Unfortunately, although our reinterpretation of the boundary states avoids
the infinities caused by non-implementable Bogoliubov transformations, it
still does not banish non-normalisable vectors completely.
(There are other ways of circumventing this problem, for example using
Connes' composition of correspondences, [27].)

In diagonalisable minimal conformal field theories the representation
space for the conformal group $G=H\times H$ decomposes into a finite
number of copies of spaces equivalent to $\hsp_V\otimes\hsp_V^*$, where
$V$ is an irreducible $\sigma$-representation of $H$ on $\hsp_V$, and $V^*$ denotes
the dual representation on the dual space $\hsp_V^*$, defined by
$V^*(x)f = f\circ V(x)^{-1} = f\circ V(x)^*$.
We may identify $\hsp_V\otimes\hsp_V^*$ with the Hilbert-Schmidt operators
${\cal L}_{HS}(\hsp_V)$ on $\hsp_V$, and the projective representations
$V$ and $\im{V}$ as the natural left and right actions on operators.
Identifying the boundary state $\Omega_\kappa$ with a linear operator it
must satisfy
$$\Omega_\kappa = V(g)\im{V}(g^\kappa)\Omega_\kappa
= V(g)\Omega_\kappa V(g)^{-1},$$
so that, by irreducibility $\Omega_\kappa$ is a multiple of the identity,
which (for infinite-dimensional $V$) is not Hilbert-Schmidt, so that
$\Omega_\kappa$ is not normalisable.

The situation is somewhat analogous to the Peter-Weyl theory for compact
groups, where the $H\times H$ representation space $L^2(H)$ decomposes
into a direct sum of ${\cal L}_{HS}(\hsp_V)$ for irreducible $V$, and the
Plancherel theorem tells us that the $\delta$ function at the identity of
$H$ is the sum of multiples of the identity in each component, except
that in this case the $V$ are finite-dimensional.
In fact, the similarity can be taken much further, if we recall that the
conformal group is a direct product group $H\times H$, with $H =
\Diff^+(S^1)$, and the subgroup $G_\kappa = \{(x,x^\kappa):x\in H\}$ is
almost a diagonal subgroup.
Were we dealing with a square-integrable representation, the fact that
$\Omega_\kappa$ is fixed by $G_\kappa$ would tell us that the projective
representation of $G$ is contained in that induced by the trivial
representation of $G_\kappa$.
(A vector $\psi$ in the representation space defines a function
$g\in G\mapsto \psi^\prime(g) = \ip{g\cdot\Omega}{\psi}$.
Since $\Omega$ is fixed by $h\in G_\kappa$, we have $\psi^\prime(gh) =
\sigma(g,h)\psi^\prime(g)$, showing that $\psi^\prime$ satisfies the
equivariance condition for the induced representation space, and for
square-integrable representations the map $\psi\mapsto \psi^\prime$ is
unitary up to a scalar factor.)
In practice this does not make sense because $G/G_\kappa \cong H$ is not
locally compact so we lack a quasi-invariant measure needed for the usual
inducing construction.
However, it formally resembles the construction of the projective
representation of $H\times H$ induced from the diagonal subgroup.
This would act on $L^2(G/G_\kappa) \cong L^2(H)$, and is the product of
the left regular $\sigma$ and right regular
$\conj{\sigma}$-representations of $H$, giving a very clear analogy with
the Peter-Weyl theory.

Fortunately Tomita theory was devised precisely to provide a remedy for
the absence of a cyclic separating vector by using only a left Hilbert
algebra.
One can still define the antilinear map $S$ as the closure of $S(a) =
a^*$, and $\Delta = S^*S$, $J = S\Delta^{-\frac12}$.
Then $\ip{b^*}{a^*} = \ip{Sb}{Sa} = \ip{a}{\Delta b}$
for a positive operator $\Delta$.
As well as the obvious relation $(ab)^* = b^*a^*$ we may set
$(ab)^* = \im{a}b^*$.
It follows that
$$\im{(ab)}c^* = [(ab)c]^* = [a(bc)]^* = \im{a}(bc)^* = \im{a}\im{b}c^*$$
and $\im{ab} = \im{a}\im{b}$, so that we have an antilinear homomoprhism
$a\mapsto \im{a}$ as before.
Moreover,
$$\im{a}c^*b^* = \im{a}(bc)^* = (abc)^* = c^*(ab)^* = c^*\im{a}b^*$$
whence $\im{a}$ commutes with $\alg_+$.
When the algebra has an identity $1 =1^*$ then
$$a^*1 = a^* = \im{a}1,$$
showing that $\Omega = 1$ is a generalised Ishibashi vector, and
we may think of the algebra as consisting of the $a\Omega$.

This is the situation in which we find ourselves in the case of the
Hilbert--Schmidt operators.
In our case with our non-normalisable state being a multiple of the
identity it is clear that we should just take the Hilbert-Schmidt
operators as the left Hilbert algebra.

In this case, since
$$\ip{b^*}{a^*} = \tr(b^{**}a^*) = \tr(a^*b) =\ip{a}{b},$$
we see that the modular operator is in this case $\Delta_V = 1$.
This means that $S_V$ is itself antiunitary, providing a slightly
different perspective on  Watts' identification of boundary states with
antiunitary maps [6].
This contrasts with the case of the free fermion model discussed
earlier, where $\Delta$ is certainly not 1.
At first sight this contradicts the fact that this is also a conformal
model.
However, those fermions were on $\real$ not the circle as in the minimal
conformal model.

One immediate consequence of the fact that $S_\kappa$ is antiunitary is
that we can extend our projective representation $U$ of the conformal
group to a unitary-antiunitary representation of the group which includes
$\kappa$, by setting $U(\kappa) = J_\kappa = S_\kappa$.
The standard Tomita-Takesaki theory tells us that
$\alg_+^\prime = J_\kappa\alg_+ J_\kappa$, and since
$\alg_+^\prime = {\alg_-}$, this shows explicitly that the quantum action
of $\kappa$ interchanges the quantum algebras of observables inside and
outside the boundary.

\sec{Multiple boundaries}

Similar methods can be applied when a region has several boundaries.
For example, when there are two boundaries associated with involutions
$\kappa_1$ and $\kappa_2$, one has to look for a state $\Omega$ which is
an eigenvector for the elements of
$G_{\kappa_1,\kappa_2} = G_{\kappa_1}\cap G_{\kappa_2}$.
This subgroup can also be thought of as $G_{\kappa_1}\cap
G_{\kappa_1\kappa_2}$, that is the subgroup of $G_{\kappa_1}$ which
commutes with the holomorphic transformation $\kappa_1\kappa_2$.
In classical conformal problems there are two common approaches to
problems in a wedge with angle $\pi/N$.
One is to calculate the Green's function using the images of the various
products of reflections in its two boundaries.
The other is to carry out the conformal transformation $z \mapsto z^N$
which maps the wedge to the half-plane where the Green's function is already
known (using a single image).
The approach we have been using shows a simple connection between these,
by producing the transformation which simplifies the problem.
In fact, the holomorphic functions invariant under $\kappa_1\kappa_2$ form
a ring of holomorphic functions of a new variable which is the transform
of $z$.

As examples we consider regions bounded by two straight lines.
There are two cases to consider, the case when the lines meet in a point,
which we take to be $0$, and the case when they are parallel in the
finite plane.
In the first case we denote by $\kappa_\theta$ the reflection in the
half-line of complex numbers with argument $\theta$.
This takes $z$ to $e^{2i\theta}\conj{z}$, and so the product
$\kappa_\theta\kappa_0$ associated with the wedge where the argument lies
in $(0,\theta)$, is the rotation which takes $z$ to $e^{2i\theta}z$.
When $\theta = \pi/N$ this rotation has finite order $N$.
By considering the Laurent expansion, any holomorphic function which is
invariant under such rotations must be a function of $z^N$.
The map $z \mapsto z^N$ is precisely the map from the wedge to the
half-plane.

The other possibility for a region bounded by two straight lines is the
strip between two parallel lines.
For definiteness let us take the strip were the imaginary part of $z$ lies
in $(0,\frac12\beta)$.
Reflection in the upper line takes $z$ to $\conj{z} + i\beta$, and the
product of the two reflections maps $z$ to $z+i\beta$.
Again we see that the holomorphic functions invariant under this
transformation are holomorphic functions of $\exp(2\pi z/\beta)$, so that
this time we have recovered the transformation $z \mapsto \exp(2\pi
z/\beta)$ which maps the strip to the half plane.

In a region with multiple boundaries one has involutions $J_{\kappa_i}$
representing the different involutions and the map $\kappa_i\kappa_j$
is represented by the linear operator $J_{\kappa_i}J_{\kappa_j}$.
In the example of the strip, double reflection of the upper half-plane
maps it to a subset of itself, and accordingly $J_{\kappa_i}J_{\kappa_j}$
gives an endomorphism of the algebra $\alg_+$.
This is very similar in form to the Longo canonical endomorphism of
$\alg_+$ defined by its image subalgebra, [17].
(That is in some ways more like the case when $\beta=0$, but with two
different algebras sharing the same boundary.)

\app{The Unruh effect in conformal field theory}

Sewell showed how to understand the Unruh effect in terms of KMS states,
[22,26].
This also shows the role of the Rindler horizon in providing a boundary
between two space-time algebras.
There is also a direct conformal field theory argument for the effect,
modelled on an argument in [4].

The world line of an observer with uniform acceleration $a$ in a fixed
direction is given in terms of the proper time $\tau$ by
$$(ct,x) = \frac{c^2}{a}(\sinh(a\tau/c),\cosh(a\tau/c)),$$
and this motivates the use of Rindler coordinates
$$(ct,x) = \frac{c^2}{a}e^{a\xi/c^2}(\sinh(a\tau/c),\cosh(a\tau/c)).$$
We may rewrite the transformation as
$$x\pm ct = \frac{c^2}{a}\exp(a(\xi\pm c\tau)/c^2).$$
This suggests, on performing a Wick rotation $t\mapsto it$, the conformal
transformation $z = c^2/a\exp(a\zeta/c^2)$, from $\zeta = \xi +ic\tau$ to
$z = x+ict$.
We calculate that $dz/d\zeta = \exp(a\zeta/c^2)$.

Suppose now that the field $\phi(z)$ has typical Fock correlation
functions
$$\expc{\phi(z_1)^*\phi(z_2)} = |z_1-z_2|^{-2h},$$
and conformal weight $(h,h)$, so that on transforming to the new
coordinates
$$\eqalign{\expc{\phi(\zeta_1)^*\phi(\zeta_2)}
&= ac^{-2}|e^{ha\zeta_1/c^2}e^{ha\zeta_2/c^2}|
\left|e^{a\zeta_1/c^2}-e^{a\zeta_2/c^2}\right|^{-2h}\cr
& = ac^{-2}
\left|e^{a(\zeta_1-\zeta_2)/c^2}-e^{-a(\zeta_1-\zeta_2)/c^2}\right|^{-2h}.\cr}
$$
Now the last expression is unchanged by the translation
$\zeta_1 \mapsto \zeta_1 + i2\pi c^2/a$,
which, in terms of the original problem involves adding $i\pi c/a$ to
$\tau_1$.
Such a periodicity in imaginary time is the KMS condition at inverse
temperature $\beta = 2\pi c/a$, thus giving the Unruh effect.

\app{The continuum Ising model}

It is known that the Ising model has a continuum limit (as the lattice
spacing goes to 0), which is described by a canonical anticommutation
relation algebra over the complex inner product space $W = {\cal
S}(\real)$, (the Schwartz functions), and one has the smeared creation
and annihilation operators
$$c(w) = \int w(x)a^*(x)\,dx, \qquad c(w)^* = \int \conj{w(x)}a(x)\,dx.$$
In the case studied in [10,16] the boundary at $x=0$ separates the
positive real axis, which is the physically interesting part of the space,
from its mirror image.
Denoting by $c(w_+)$ the operator which creates the fermion state $w_+$ on
the physically interesting side of the boundary, it turns out that the
boundary state $\Omega_\kappa$ satisfies
$c(w_+)^*\Omega_\kappa = c(Kw_+)\Omega_\kappa$, for a certain operator
$K$, mostly simply expressed in terms of the Fourier transform
${\cal F}W_+(p)$ by
$({\cal F}Kw_+)(p) = K(p)\conj{({\cal F}w_+)(-p)}$,
where $K(p) = -ip/(E_p\pm m)$, $E_p = \sqrt{p^2+m^2}$ and the sign depends
on the type of boundary condition.
([10] expresses $K$ in terms of the rapidity $\theta$ rather than
$p=m\tanh\theta$.)
This fits our previous framework with $\alpha_\kappa(c(w_+)) = c(Kw_+)$.

Subtleties arise because the algebra is represented on the standard
Fock--Dirac space generated by a vacuum killed by creators of negative
energy states and by annihilators of positive energy states, so that
there is a second boundary in momentum space, the Fermi level of the free
Dirac theory.
Here the boundary separates positive from negative energies, and, as
above, the Dirac vacuum state $\Omega$ is killed by creators of negative
energy states and by annihilators of positive energy states.
The annihilators of negative energy states $c(w_-)^*$ are then
reinterpreted as creators of a positron $\im{c}({\cal C}w_-)$ (${\cal C}$
being charge conjugation), and the defining identity
$c(w_-)^*\Omega = \im{c}({\cal C}w_-)\Omega$ can be interpreted as another
example of the same class.

As mentioned in Section 3, [10,16] try to interpret $K$ in terms of a
Bogoliubov transformation with $K = Z= BA^{-1}$.
However, the operator $K^*K$ can be considered as the integral operator
with the distributional kernel $k(p,q)=|K(p)|^2\delta(p-q)$, from which it is
obvious that the Hilbert--Schmidt norm
$$\tr(K^*K) = \int_\real k(p,p)\,dp$$
diverges, and so the Bogoliubov transformation is not implementable.

Since $K$ is normal, the condition $A^*A+B^*B = 1$ reduces to
$(AA^*)^{-1} = 1+K^*K$, so we take
$$A = (1+K^*K)^{-\frac12}:W_+ \to W_+ \qquad{\rm and}\qquad
B = KA:W_+ \to W_-.$$
The adjoint antilinear map $K^*:W_- \to W_+$ can be similarly used to
extend the operators $A$ and $B$ to $W_-$ by defining
$$A = (1+K^*K)^{-\frac12}:W_- \to W_- \qquad{\rm and}\qquad
B = -K^*A:W_- \to W_+.$$
(It can  be shown that this choice is essentially unique, [21].)
Then for $w\in W$ we set $c_K(w) = c(Aw) - c(Bw)^*$.

The Fock vacuum vector $\Omega_\kappa$, killed by the annhilation
operators $c_K(w)^*$, therefore satisfies
$$0 = c_K(A^{-1}w_+)^*\Omega_\kappa
= c(w_+)^*\Omega_\kappa -c(Kw_+)\Omega_\kappa,$$
giving the required image condition on $\Omega_\kappa$.
(There is a second condition that
$$0 = c_K(A^{-1}w_-)^*\Omega_\kappa
= c(w_-)^*\Omega_\kappa + c(K^*w_-)\Omega_\kappa,$$
from which we deduce that
$c(w_-)^*\Omega_\kappa = -c(K^*w_-)\Omega_\kappa$
for all $w_-\in W_-$.)

As mentioned in Section 3, the restriction to $\car(W_+)$ of the state
defined by  $\Omega_\kappa$ is quasi-free.
The injection of $W_+$ into the double $W$ is given by
$I_Kw_+ = Aw_+ + Bw_-$.

It is now easy to compute $S$ at the one particle level, since we have
$$Sc(w_+)\Omega_\kappa  = c(w_+)^*\Omega_\kappa = c(Kw_+)\Omega_\kappa$$
and, since $S$ is an involution,
$$Sc(w_-)^*\Omega_\kappa = c(K^{-1}w_-)\Omega_\kappa.$$
Similarly, we have
$$Sc(w_-)\Omega_\kappa  = -c(K^*w_+)\Omega_\kappa,
\qquad
Sc(w_+)^*\Omega_\kappa = -c({K^*}^{-1}w_+)\Omega_\kappa.$$

Thus on the one-particle space $S$ has the matrix form
$$S \sim \left(\matrix{0 &K^{-1} &0 &0\cr K &0 &0 &0\cr
0 &0 &0 &-{K^*}^{-1}\cr 0 &0 &-K^* &0\cr}\right),$$
giving
$$\Delta \sim S^*S
= \left(\matrix{K^*K &0 &0 &0\cr 0 &(KK^*)^{-1} &0 &0\cr
0 &0 &KK^* &0 \cr 0 &0 &0 &(K^*K)^{-1}\cr}\right).$$
Thus in this case there is a non-trivial modular operator, and $S$ is not
antiunitary.

It is well-known that the KMS condition facilitates the calculation of
correlation functions.
For example, we have
$$\eqalign{\expc{c(w_+)^*c(z_+)}  &= \expc{c(z_+)\Delta c(w_+)^*\Delta}\cr
&= \expc{c(z_+)c(K^*Kw_+)^*}\cr
&= \ip{K^*Kw_+}{z_+} - \expc{c(K^*Kw_+)^*c(z_+)},\cr}$$
so that
$\expc{c((1+K^*K)w_+)^*c(z_+)}  = \ip{K^*Kw_+}{z_+}$,
and
$$\expc{c(w_+)^*c(z_+)}  = \ip{(1+K^*K)^{-1}K^*Kw_+}{z_+}.$$
This illustrates the fact that this interpretation of the boundary states
also provides a useful tool for calculation.

\bigskip\noindent
{\bf Acknowledgements}

The second author is grateful to INDAM and to EPSRC who funded parts of
this work.

\bigskip\noindent
{\bf References}

\medskip
\ref{A} H Araki, \lq\lq On quasifree states of CAR and Bogoliubov
automorphisms\rq\rq, {\it Publ. Res. Inst. Math. Sci.} {\bf 6} (1970/71),
385-442.
\ref{FB} L Birke and J Fr\"ohlich,\lq\lq KMS etc.\rq\rq, {\it Rev. Math.
Phys.} {\bf 14} (2002), 829-871.
\ref{C} JL Cardy, \lq\lq Effect of boundary conditions on the operator
content of two-dimensional conformally invariant theories\rq\rq, {\it Nuclear
Phys. B} {\bf 275} (1986) 200-218.
\ref{C1} JL Cardy, \lq\lq Conformal invariance\rq\rq, in {\it Phase
transitions and critical phenomena} {\bf 11} ed C Domb and JL Lebowitz,
Academic Press, London 1987.
\ref{C2} JL Cardy, \lq\lq Boundary conditions, fusion rules and the
Verlinde
formula\rq\rq, {\it Nuclear Phys. B} {\bf 324} (1989) 581.
\ref{C+L} JL Cardy and  DC Lewellen, \lq\lq Bulk and boundary operators in
conformal field theory\rq\rq, {\it Phys. Lett. B} {\bf 259} (1991),
274-278.
\ref{Co} A Connes, {\it Noncommutative Geometry}, Academic Press, San
Diego, 1994.
\ref{vD} A van Daele, \lq\lq A new approach to the Tomita--Takesaki theory
of generalised Hilbert algebras\rq\rq, J. Func. Analysis, {\bf 15} (1974),
378-393.
\ref{E+K} DE Evans and Y Kawahigashi, {\it Quantum symmetries on operator
algebras}, Clarendon Press, Oxford, 1998.
\ref{G+Z} S Ghoshal and A Zamolodchikov, \lq\lq Boundary S matrix and
boundary states in two-dimensional integrable quantum field theory\rq\rq,
{\it Int. J. Mod. Phys. A} {\bf 9} (1994), 3841-3885, hep-th/9306002.
\ref{GJ} J Glimm and A Jaffe, {\it Quantum Physics}, Springer Verlag, New
York, 1981.
\ref{GQS} V Guillemin, D Quillen, and S Sternberg, \lq\lq The
classification of the irreducible complex algebras of infinite type\rq\rq,
{\it J. Analyse Math.} {\bf 18} (1967), 107-112.
\ref{Im} C Isham private communication.
\ref{I} N Ishibashi, \lq\lq The boundary and crosscap states in conformal
field theories\rq\rq, {\it Mod. Phys. Lett. A} {\bf 4} (1989), 251-264.
\ref{K+R} RV Kadison and JR Ringrose {\it Fundmentals of the theory of
operator algebras, Vol II}, Academic Press, London, 1986.
\ref{LMSS} A Leclair G Mussardo H Saleur and S Skorik, \lq\lq Boundary
energy and boundary states in integrable quantum field theories\rq\rq,
{\it Nucl. Phys. B} {\bf 453} (1995), 581-618, hep-th/9503227.
\ref{L} R Longo, \lq\lq Simple injective subfactors\rq\rq, {\it Adv. in
Math.} {\bf 63} (1987), 152-171.
\ref{OS} K Osterwalder and R Schrader, \lq\lq Axioms for Euclidean Green's
functions, I, II\rq\rq, {\it Commun. Math. Phys.} {\bf 31} (1973) 83-112,
and {\bf 42} (1975) 281-305.
\ref{PS} R Powers and E St\o rmer, \lq\lq Free states of the canonical
anticommutation relations\rq\rq, {\it Commun. Math. Phys.} {\bf 16}
(1970), 1-33.
\ref{PR} RJ Plymen and PL Robinson, {\it Spinors in Hilbert space},
Cambridge
tracts in mathematics {\bf 114}, Cambridge University Press, 1994.
\ref{MS} M Semplice {\it Boundary Conformal Fields and Tomita--Takesaki
theory}, DPhil thesis, Oxford 2002.
\ref{S} G Sewell, \lq\lq Quantum fields on manifolds: PCT and
gravitationally induced thermal states\rq\rq, {\it Ann. of Phys.}
{\bf 141} (1982), 201-224.
\ref{St} S Stratila and L Zsido, {\it Lectures on von Neumann algebras},
(English edition), Abacus Press, Tunbridge Wells, 1979.
\ref{T} M Takesaki, {\it Tomita's theory of modular Hilbert algebras and
its applications}, Lecture Notes in Mathematics {\bf 128}, Springer Verlag,
Heidelberg, 1970
\ref{U} WG Unruh, \lq\lq Notes on black hole evaporation\rq\rq, {Phys.
Rev. D} {\bf 14} (1976), 870.
\ref{Wa} RM Wald, {\it Quantum field theory in curved spacetime and black
hole thermodynamics}, University of Chicago Press, Chicago, 1994.
\ref{W} AS Wassermann, Seminar, Oxford 2002.

\bye